\documentclass{article}
\usepackage{spconf,amsmath,graphicx}
\usepackage{subcaption}
\usepackage{amsfonts}
\usepackage{amssymb}
\usepackage{mathrsfs}

\title{The Importance of the Instantaneous Phase \\ for classification using Convolutional Neural Networks}
\name{Luis Sanchez Tapia, Marios S. Pattichis, Sylvia Celedon-Pattichis, Carlos Lopez Leiva}
\address{University of New Mexico\\
	Electrical and Computer Engineering\\
	Albuquerque, New Mexico}

\begin{document}
%
\maketitle
\begin{abstract}

Large scale training of Convolutional Neural Networks (CNN) is extremely demanding in terms of computational resources. 
Also, for specific applications, 
the standard use of transfer learning also tends to require far
more resources than what may be needed.

This work examines the impact of using AM-FM representations as input images for CNN classification applications. A comparison was made between AM-FM components combinations and gray scale images as inputs for reduced and complete networks. 

The results showed that only the phase component produced significant predictions within a simple network. Neither IA or gray scale image were able to induce any learning in the system. Furthermore, the FM results were 7x faster during training and used 123x less parameters compared to state-of-the-art MobileNetV2 architecture, while maintaining comparable performance (AUC of 0.78 vs 0.79).

\end{abstract}
\section{Introduction}
\label{sec:intro}

Convolutional Neural Networks dominates image and video analysis methods. Their ever growing utilization had raise the complexity to the point they are understood as a black box. These deeper networks consist of parameters in the order of the millions, therefore transforming the training process into a long and less intuitive procedure. In order to avoid over fiting, large amounts of ground truth samples needs to be generated to train even a small system. As a response, transfer learning methods were developed to re use trained layers into common basic patterns between applications.

A generalize practice among the CNN implementation is the use of regular gray scale images and instead, focus the innovation towards architecture design. To answer that tendency, this work aims to explore the relevance of amplitude and phase representations as input images for CNN applications. Motivated by earlier research regarding the phase information \cite{oppenheim} and the development of recent AM-FM representations \cite{multiscale}, the importance of the instantaneous phase in images needs to be investigated in the context of the current explosive use of CNNs. 

We are going to pick a classic application in face recognition to explore the impact of the AM-FM representation. A collection of videos was selected from the Advancing Out-of-school Learning in Mathematics and Engineering (AOLME) \cite{aolme1}  after-school program. This dataset is suitable for studying real-life unconstrained environments as it resembles the actual hard problems machine learning methods are facing.

The contributions includes the exploration of the Amplitude-Modulation Frequency-Modulation (AM-FM) components as inputs for low-parameter CNNs. An examination for the performance using the Instantaneous Amplitude, Instantaneous Phase and the full AM-FM representation as input images compared to regular gray scale images. We provide a transfer learning approach by using a fixed filterbank as the lower level of the network, this aims to reduce training overhead and increase the interpretability of the model. This Gabor filterbank was designed for implementing Dominant Component Analysis using small size kernels for hardware implementation. Also, an exploration in reduction of actual CNN architectures to investigate the limits of the AM-FM components for image classification.

Our results demonstrates that FM representations allow for a very fast convergence and training in reduced simple networks, while matching the performance of much more complex network using regular gray scale images. 

The paper is organized as follows, background section depicts the dataset generation and other work with AM-FM. The Methodology section describes the low-parameter AM-FM filterbank design and the low-parameter regression architecture for detecting faces inside blocks. Next, the result section provides a comparison among AM-FM components and gray scale input images. Finally, the conclusion and future work.

\section{Background}
\label{sec:background}


\subsection{Previous AM-FM Research}

Previous research had establish the multiscale AM-FM estimation using Dominant Component Analysis (DCA) to be the most popular method for AM-FM represenations. 
The AM-FM methods developed in \cite{medical3} and \cite{medical17} provided the basis for the DCA methods used in this research. In contrast to previous AM-FM implementations, the AM-FM filters were designed to be as small as possible to allow fast implementation within large datasets. Therefore, allowing a chance to compete against unprocessed gray scale images.

A previous example of research with AOLME dataset and AM-FM representation is presented in \cite{wenjingshi}. The authors extract AM-FM features to identify regions of interest with the goal of detecting the front and back heads. Then, further processing would determine where students are focused their attention. In \cite{shi3} the same authors used AM-FM methods to determine student group interactions such as human to human, human to others and no interactions. 

The current research extends the work by \cite{wenjingshi} in the following directions: a small kernel filterbank of $ 11\times11$ compared $21\times21$ to provide 4 times less trainable parameters per filterbank. In addition the reduction from 6 scales to 2 scales further reduced the computation time required and the amount of parameters to tune. This reduced filterbank would allow future hardware reconfigurable implementations similar to \cite{llamocca}. Another extension consists in a much larger dataset based on image blocks, within a longer span of time and different angles of the camera at each video. An most importantly, a comparison is provided for the combinations of AM-FM representations versus raw gray scale images for block based face detection. 

\subsection{Deep Learning}
Two extremes are explored in this work. A reduced network based on the very well tested LeNet-5 by Yann LeCun \cite{lecun98}. With only one convolutional layer and two fully connected, this network is suitable for proofs of concepts experiments regarding processing over input images. In the other hand, MobileNetV2 \cite{mobilenetv2} represents the state of the art in low complexity neural nets with low power consumption for embedded devices. MobileNet V2 is considerably faster than other CNN architectures, since it was specifically targeted towards mobile applications.

\section{Methodology}
\label{sec:methodology}

The general workflow consists of frames being processed by Dominant Component Analysis and cashed into memory for later use. Then, a selection of IA, FM or IA+FM components are fed to the Single-Block regression system to produce block predictions.

\subsection{Gabor Filterbank}

\begin{figure}
	\centering
	\begin{subfigure}{0.5\columnwidth}
		\centering
		\includegraphics[width=0.9\columnwidth]{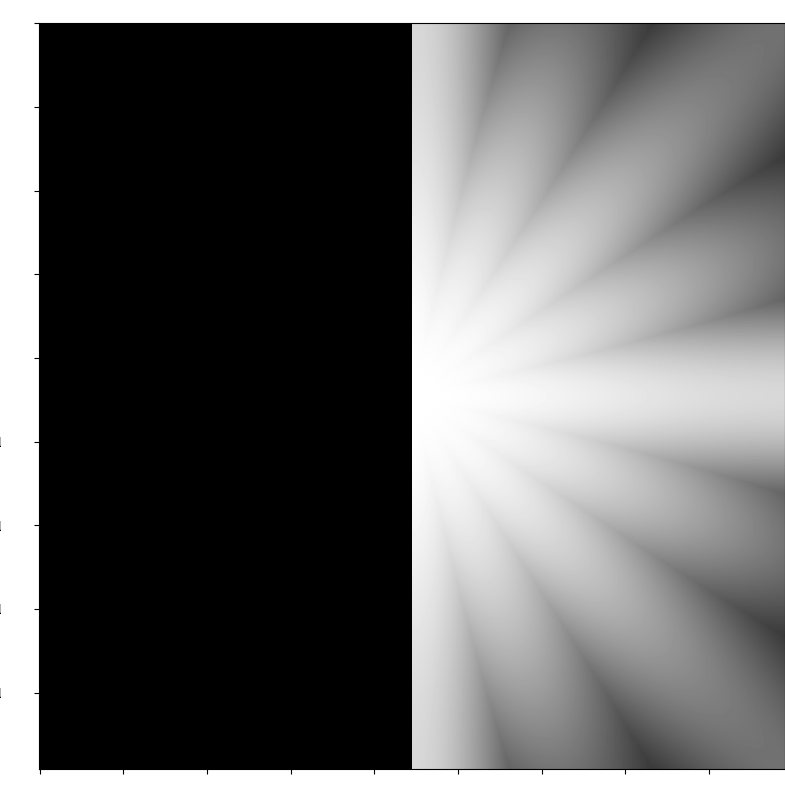}
		\caption{Directional Gaussian}
		\label{fig:FFTGaussian}
	\end{subfigure}%
	\begin{subfigure}{0.5\columnwidth}
		\centering
		\includegraphics[width=0.9\columnwidth]{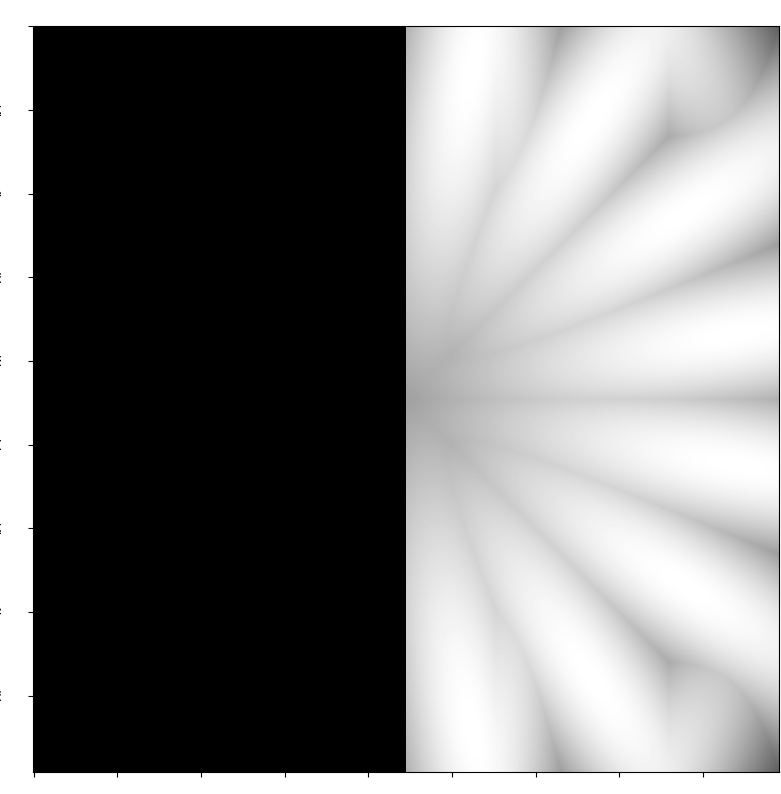}
		\caption{Directional Gabor}
		\label{fig:FFTGabor}
	\end{subfigure}
	\caption{FFT magnitude plot from Low-parameter Filterbank implementations using $ 11 \times 11 $ filter coefficients.}
	\label{fig:FFTMAX}
\end{figure}

To generate the Gabor filterbank used to estimate the AM-FM components, a directional Gaussian was rotated 8 times using a step of $ \theta += 0.39 $ radians each time. Then, another set of directional Gabor were distributed to cover the whole frequency spectrum. The directional shape was obtained with $ \sigma_{x} = 1.5 $ and $ \sigma_{y} = 1.5/4 $. The following figures depict the FFT magnitude plots of the low-parameter Directional Gabor filterbank. In Figure \ref{fig:FFTMAX}: subplot \ref{fig:FFTGaussian} shows the FFT Magnitude for directional Gaussian, subplot \ref{fig:FFTGabor} shows the FFT magnitude of the second scale of the Gabor filterbank.

\begin{figure}	[bt]
	\includegraphics[width=\columnwidth]{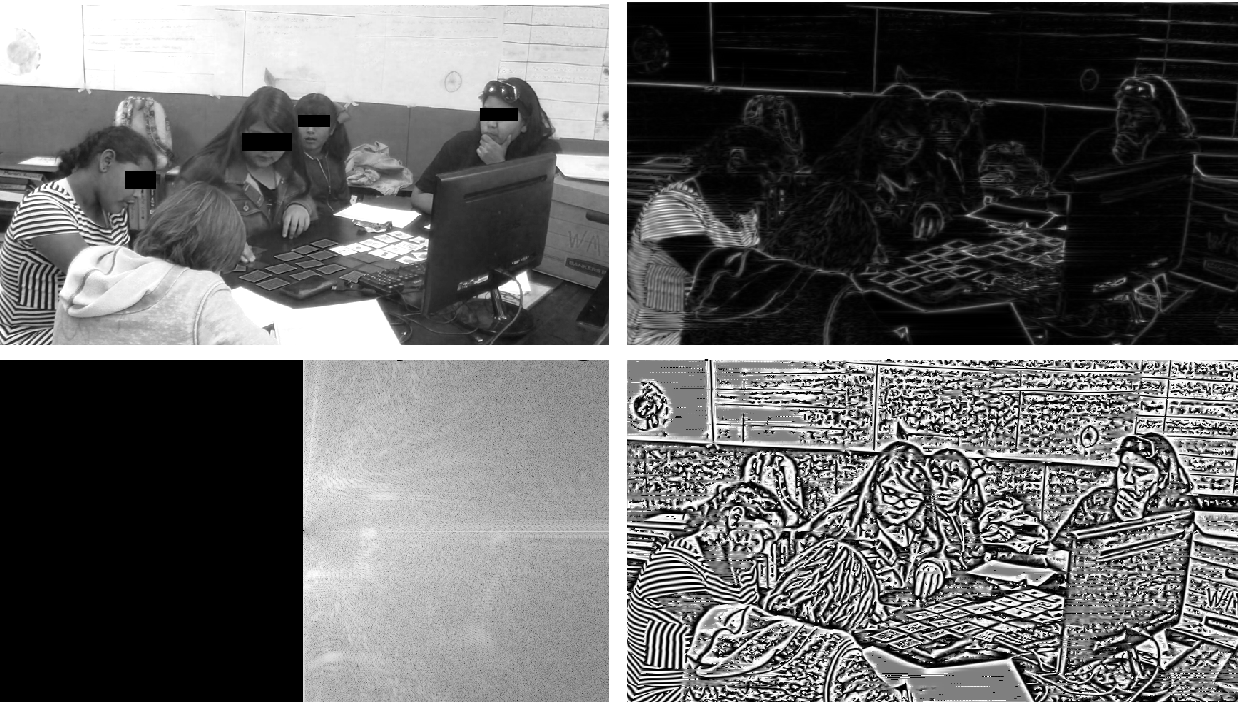}
	\caption{Top-left: Original image. Bottom-left: FFT Magnitude of analytical image. Top-right: IA from AM-FM filterbank using DCA. Bottom-right: FM component using DCA. Undesirable AM components can be seen at the lower-left corner of the IA image.}
	\label{fig:CosPhiHilbert}
\end{figure}

\subsection{Reduced-LeNet Block-based Architecture}

The reduced-LeNet architecture started as an exploration of the impact that the type of input data have over the effectiveness of a CNN architecture. LeNet-5, despite it's simplicity, has proven to be an effective baseline architecture for regression based on the dominant FM component. Starting from the baseline architecture, hyper-parameter tuning was used to reduce over-fitting. 
Without sacrificing performance, the number of convolutional layers was reduced from 3 to 1, followed by an increased Max Pooling size of $ 5 \times 5$. The final architecture is given in table \ref{tab:reducedlenetTable}. 

\begin{table}[bt]
	\centering
	
	\begin{center}
	\begin{tabular}{llllll}
		\textbf{Layer} & \textbf{Type} & \textbf{Kernel} & \textbf{Size} & \textbf{Stride} & \textbf{Act.} \\

		In             & Input         & -               & 50x50         & -               & -             \\
		C1             & Conv2D        & 6 (5x5)         & 46x46         & 1               & selu          \\
		S2             & MaxPool       & 6 (5x5)         & 23x23         & 2               & -             \\
		F3             & Dense         & -               & 40            & -               & selu          \\
		F4             & Dense         & -               & 24            & -               & selu          \\
		Out            & Dense         & -               & 1             & -               & sigmoid      
	\end{tabular}
	\end{center}
	
	\caption{ Block-based regression architecture for face detection. The input block is of size $ 50 \times 50 $.}
	\label{tab:reducedlenetTable}
\end{table}

\section{Results}
\label{sec:results}

The results of the reduced-LeNet for single block regression architectures were compared for: (i) Original gray scale image, (ii) DCA estimates for the FM image, (iii) DCA estimates for the IA image and (iv) IA and FM components combined as different channels. Additionally, gray scale images and FM images were tested using the MobileNetV2 architecture. Several thresholds were tested in correspondence with the range values of the output predictions.

For this work, the training set consists of 12 video clips extracted from 12 different sessions. The test set consists of 6 video clips extracted from 6 other sessions, so there is no chance of mixing these 2 sets. 
At each video clip, one frame is extracted for each minute, resulting in 24 frames per video. Each frame is reduced to half size and then zero padded to extend the size to the nearest multiple of 5 for the rows and 9 for the columns, giving a total of 45 blocks of the same size: $ 50 \times 50 $ pixels. Finally, 12960 blocks were extracted to train and validate the network, while 6480 blocks from different frames were used for testing.

The selection of the Receiver Operating Characteristic (ROC) and the Area Under the Curve (AUC) as metrics involves the unbalanced nature of the dataset, since the blocks with a face percentage comprises around 10\% of the total of blocks in the frame.

\subsection{Face-Detection using Single-Block Classification}

\begin{figure}	[bt]
	\includegraphics[width=\columnwidth]{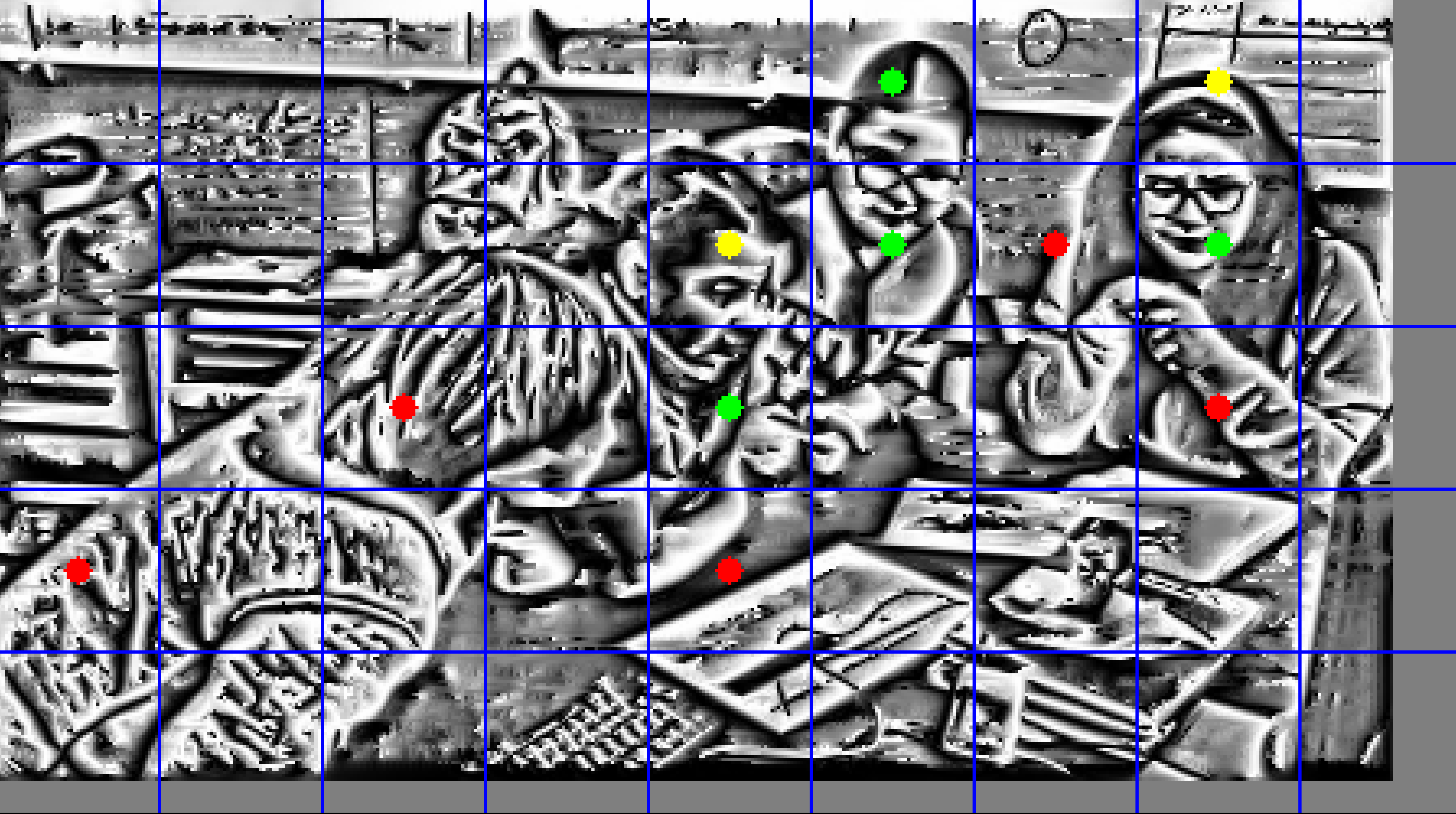}
	\caption{Single-Block regression: FM frame with marks for True Positives (green), False Positives (red) and False Negatives (yellow).}
	\label{fig:frame_lenet_cosPhi}
\end{figure}

Several frames were used for debugging the reduced-LeNet block-based face detection. 
The original image in gray scale showed no face predictions in any of the 45 blocks in the frame. Same behavior is encountered in with the AM plus FM image, regardless of the given threshold value. The results from the FM input image demonstrated that the proposed method managed to find 2 faces along with several False Positives (red marks), presented in figure \ref{fig:frame_lenet_cosPhi}. 

The following figure \ref{fig:roc_cosPhi_original} demonstrate the importance of the phase component in block-based face detection using CNNs. The ROC for the original grayscale image describes a random predictor, not capable of differentiating between the features of the images. In contrast, the phase does give a functional predictor to start exploring the face detection methods. Similar results are obtained for the AM components against the FM components alone. For both IA and FM components, the results were similar to the ones obtained by the FM components.

\begin{figure}	[bt]
	\includegraphics[width=\columnwidth]{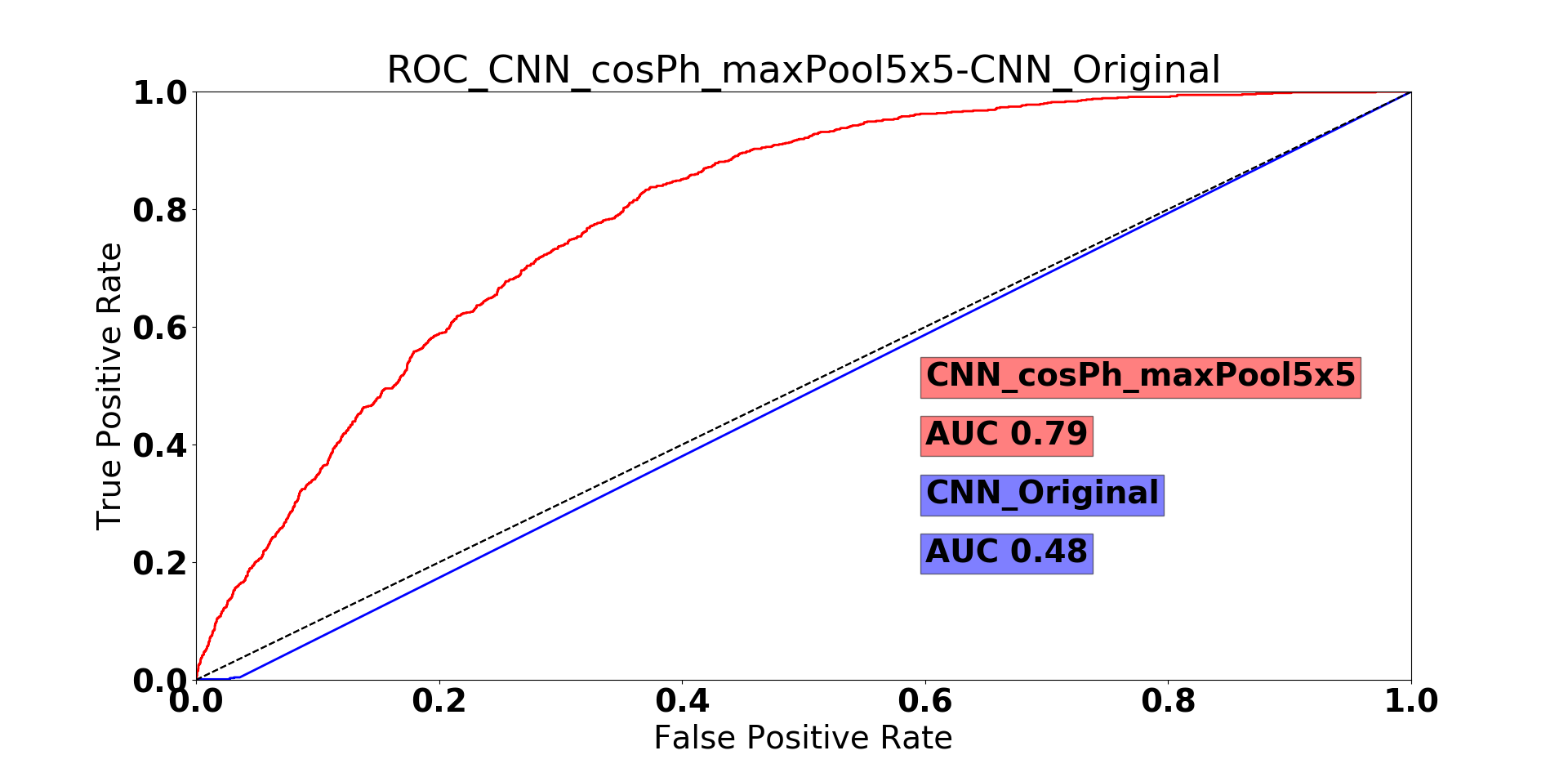}
	\caption{Single-Block: ROC curve for FM component (in red) and original image (in blue).}
	\label{fig:roc_cosPhi_original}
\end{figure}

\section{Results for MobileNetV2 }
\begin{figure}	[bt]
	\includegraphics[width=\columnwidth]{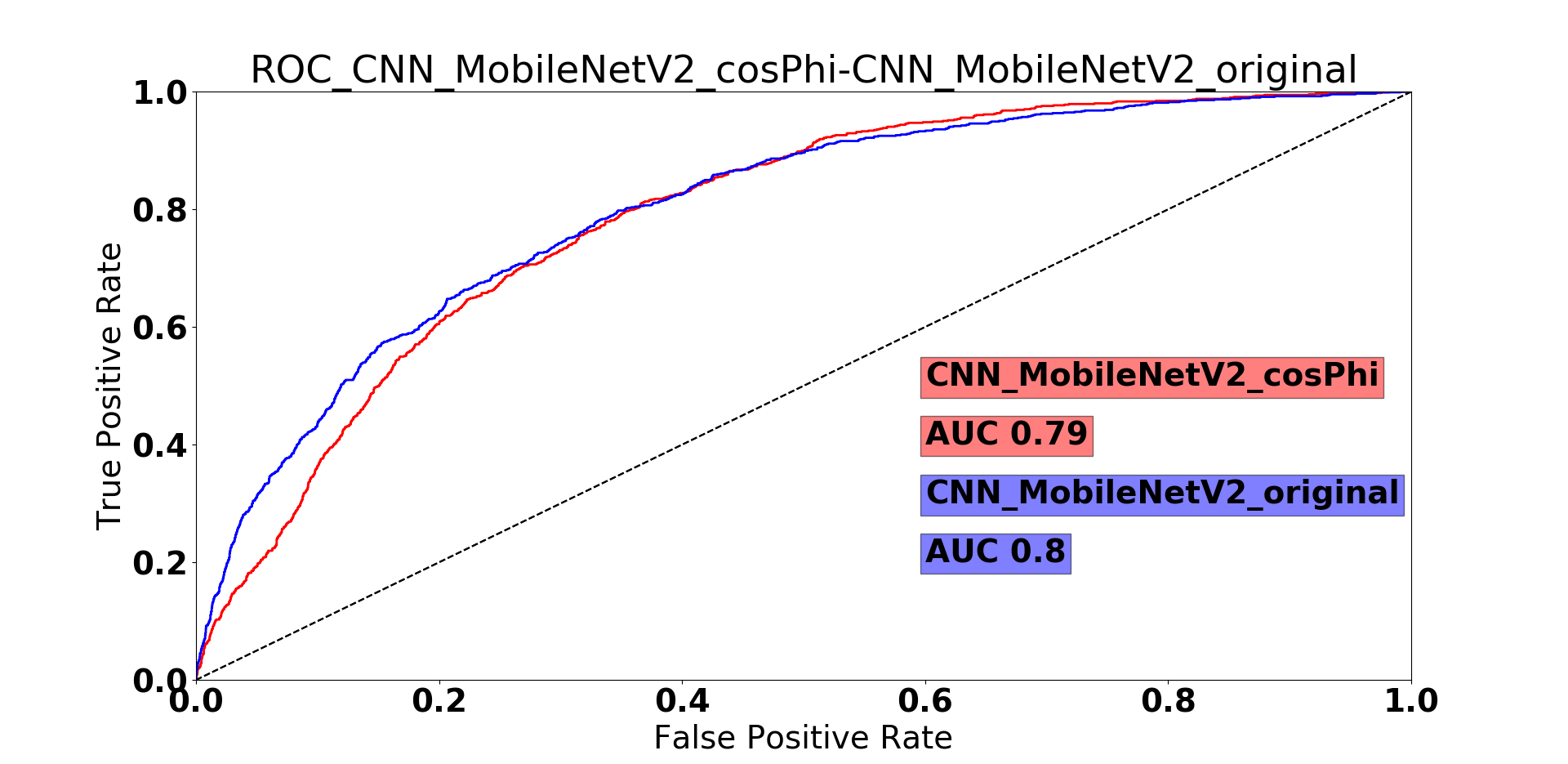}
	\caption{MobileNet V2 regression: ROC curve for FM component (in red) and original image (in blue).}
	\label{fig:roc_mobile_cosPhi}
\end{figure}

MobileNetV2 network was used by importing the model as layer and set the input-output layers to train our dataset. The ROC in figure \ref{fig:roc_mobile_cosPhi} demonstrates the similarities between training with the original image and the FM component. The AUC values are also similar. However, the required training time for MobileNet V2 is significantly more than for the proposed reduced-LeNet. From table \ref{tab:mobileProposedTable}, it is clear that the proposed architectures are 7x to 11x faster to train. Furthermore, in terms of generalization, the proposed architecture use 123 times less trainable parameters than MobileNet V2.

\begin{table}[bt]
	\centering
	
	\begin{center}
		\begin{tabular}{l l l l} 
			\textbf{Network}      & \textbf{GTX 1050} & \textbf{Two GTX 1080} & \textbf{Param.}  \\ [0.5ex] 

			MobileNet V2 &220 sec           & 68 sec        & 1205073 \\ 

			Reduced LeNet &20  sec         & 10  sec       & 9775  \\ 

		\end{tabular}
	\end{center}
	
	\caption{Comparison of MobileNet V2 vs the proposed architectures. The times correspond to a single epoch. For the laptop, we used Intel core i5-7300HQ CPU @ 2.50GHz, 8GB RAM and NVIDIA GeForce GTX 1050. For the desktop system, we used an Intel Xeon CPU ES-2630 v4 @ 2.20GHz, 32 GB RAM and two NVIDIA GeForce GTX 1080. The GPU was used during training. }
	\label{tab:mobileProposedTable}
\end{table}


\section{Conclusions and Future Work}
\label{sec:conlusions}
This work has demonstrated the importance of using the dominant Frequency-Modulation components in the training of Convolutional Neural Networks in block-based face detection applications. The results from this study suggest that using the FM component of an image provides significantly better results and much faster training with low-parameter networks. Reduced-LeNet, a vastly simplified version of LeNet-5 with a single convolutional layer achieved an $ AUC = 0.78 $ compared to $AUC = 0.48$ for the original image as input. In comparison, the MobileNet V2 network achieved an $AUC = 0.8$ with the original image input, while requiring 11x more time to train and 123x more parameters to train. 

Future work include the further exploration of methods that used FM components as input. Future research should consider the replacement of the block architecture by region of interest methods. Fixed-point implementation on FPGAs can also lead to significant speed-up in training.


\section{Acknowledgments}
\label{sec:Acknowledgments}

This material is based upon work supported by the National Science Foundation under Grant No. 1613637 and Grant No. CNS-1422031. Any opinions or findings of this thesis reflect the views of the author. They do not necessarily reflect the views of NSF.

\bibliographystyle{IEEEbib}
\bibliography{references_Luis}

\begin{thebibliography}{1}

\bibitem{oppenheim}
A.~V. {Oppenheim} and J.~S. {Lim},
\newblock ``The importance of phase in signals,''
\newblock {\em Proceedings of the IEEE}, vol. 69, no. 5, pp. 529--541, May
  1981.

\bibitem{multiscale}
V.~{Murray}, P.~{Rodriguez}, and M.~S. {Pattichis},
\newblock ``Multiscale am-fm demodulation and image reconstruction methods with
  improved accuracy,''
\newblock {\em IEEE Transactions on Image Processing}, vol. 19, no. 5, pp.
  1138--1152, 2010.

\bibitem{aolme1}
Sylvia Celedon-Pattichis, Carlos~Alfonso LopezLeiva, Marios~S. Pattichis, and
  Daniel Llamocca,
\newblock ``An interdisciplinary collaboration between computer engineering and
  mathematics/bilingual education to develop a curriculum for underrepresented
  middle school students,''
\newblock {\em Cultural Studies of Science Education}, vol. 8, no. 4, pp.
  873--887, Dec 2013.

\bibitem{medical3}
C.~Agurto, V.~Murray, H.~Yu, J.~Wigdahl, M.~Pattichis, S.~Nemeth, S.~Barriga,
  and P.~Soliz,
\newblock ``A multiscale optimization approach to detect exudates in the
  macula,''
\newblock {\em IEEE Journal of Biomedical and Health Informatics}, vol. 18, no.
  4, pp. 1328--1336, July 2014.

\bibitem{wenjingshi}
WENJING SHI,
\newblock ``Human attention detection using am-fm representations,''
\newblock M.S. thesis, University of New Mexico, 2016.

\bibitem{shi3}
W.~{Shi}, M.~S. {Pattichis}, S.~{Celedón-Pattichis}, and C.~{LópezLeiva},
\newblock ``Dynamic group interactions in collaborative learning videos,''
\newblock in {\em 2018 52nd Asilomar Conference on Signals, Systems, and
  Computers}, Oct 2018, pp. 1528--1531.

\bibitem{llamocca}
Daniel Llamocca and Marios Pattichis,
\newblock ``A self-reconfigurable platform for the implementation of 2d
  filterbanks with real and complex-valued inputs, outputs, and filter
  coefficients,''
\newblock {\em VLSI Design}, vol. 2014, 05 2014.

\bibitem{lecun98}
Y.~LeCunn, L.~Bottou, Y.~Bengio, and P.~Haffner,
\newblock ``Gradient-based learning applied to document recognition,''
\newblock {\em Proceedings of the IEEE}, vol. 86, no. 11, pp. 2278--2324,
  November 1998.

\bibitem{mobilenetv2}
Mark Sandler, Andrew~G. Howard, Menglong Zhu, Andrey Zhmoginov, and
  Liang{-}Chieh Chen,
\newblock ``Inverted residuals and linear bottlenecks: Mobile networks for
  classification, detection and segmentation,''
\newblock {\em CoRR}, vol. abs/1801.04381, 2018.

\end{thebibliography}

\end{document}